\documentclass[%
 reprint,
superscriptaddress,
 amsmath,amssymb,
 aps,
prl
]{revtex4-2}

\usepackage{graphicx}
\usepackage{dcolumn}
\usepackage{bm}
\usepackage{ulem}
\usepackage{placeins}

\usepackage{lipsum}
\usepackage{color}
\usepackage{arydshln}

\begin{document}

\preprint{APS/123-QED}

\title{Contact line catch up by growing ice crystals}




\author{Rodolphe Grivet}
\email{rodolphe.grivet@ladhyx.polytechnique.fr}
\affiliation{Laboratoire d'Hydrodynamique (LadHyX), UMR 7646 CNRS-Ecole Polytechnique, IP Paris, 91128 Palaiseau CEDEX, France}%

\author{Antoine Monier}
\affiliation{Sorbonne Universit\'e, CNRS, UMR 7190, Institut Jean Le Rond $\partial$'Alembert,  F-75005 \ Paris, France}%

\author{Axel Huerre}%
 \affiliation{MSC, Universit\'e de Paris, CNRS (UMR 7057), 75013 Paris, France}%
 
 \author{Christophe Josserand}
\affiliation{Laboratoire d'Hydrodynamique (LadHyX), UMR 7646 CNRS-Ecole Polytechnique, IP Paris, 91128 Palaiseau CEDEX, France}%

\author{Thomas S\'eon}
\affiliation{Sorbonne Universit\'e, CNRS, UMR 7190, Institut Jean Le Rond $\partial$'Alembert,  F-75005 \ Paris, France}%

\date{\today}

\begin{abstract}

The effect of freezing on contact line motion is a scientific challenge in the understanding of the solidification of capillary flows. 
In this letter, we experimentally investigate the spreading and freezing of a water droplet on a cold substrate. 
We demonstrate that solidification stops the spreading because the ice crystals catch up with the advancing contact line. Indeed, we observe the formation and growth of ice crystals along the substrate during the drop spreading, and show that their velocity equals the contact line velocity when the drop stops. 
Modelling the growth of the crystals, we predict the shape of the crystal front and show that the substrate thermal properties play a major role on the frozen drop radius. 

\end{abstract}

\maketitle


When freezing occurs during capillary flows, such as for droplets impacting very cold substrates \cite{thievenaz2020freezing,gielen2020solidification} or trickles flowing over subzero substrates \cite{monier2020freezing,huerre2021solidification}, the modification of wetting due an ice layer formation leads to the creation of surprising ice patterns.
The interplay between contact line motion and phase change received increasing attention over the past years \cite{deegan1997capillary, antonini2013water, jambon2018singular}. In particular, understanding the dynamics of contact line in the presence of solidification enables the tuning of the interaction between the ice and the solid \cite{amirfazli2016fundamentals, kreder2016design} and finds numerous applications over various fields going from metallurgy \cite{cieslak1991welding} to icing/de-icing in aeronautics~\cite{cebeci2003aircraft} or inkjet like ice printing \cite{zheng2020inkjet}. 

An important question that arises often in these situations, is the arrest of a moving contact line because of solidification. 
Despite numerous studies looking at the arrest criteria for a moving contact line on a sub-cooled substrate, the different potential mechanisms are still debated, lacking a global understanding.
They include dissipation increase at the contact line due to solidification \cite{schiaffino1997molten,herbaut2020criterion}, solidification of a critical volume \cite{tavakoli2014spreading}, or crystal growth pinning \cite{de2017contact,herbaut2019liquid,koldeweij2021initial}.

In this letter, we investigate experimentally the spreading and arrest of a water drop deposited on a cooled substrate for temperatures down to $T_{\rm s} = -25^{\circ} \mathrm{C}$. We observe that, in this temperature range, the colder the substrate, the smaller the arrest radius. 
We demonstrate for the first time that this is due to ice crystals that catch up with the advancing contact line.
Indeed, we observe the ice crystals growing in the drop at the contact with the substrate while the droplet spreads, and show that the crystal growth velocity along the substrate equals the contact line advancing velocity at the time of arrest. 
Based on the strong heat transfer between the ice and the substrate, we propose a model for the crystal growth dynamics that matches our experimental measurements and allows us to predict the shape of the crystal front. Our study discriminates therefore between the debated mechanisms behind contact line arrest due to solidification. 

\begin{figure}[h!]
\centering
\includegraphics[width=0.45\textwidth]{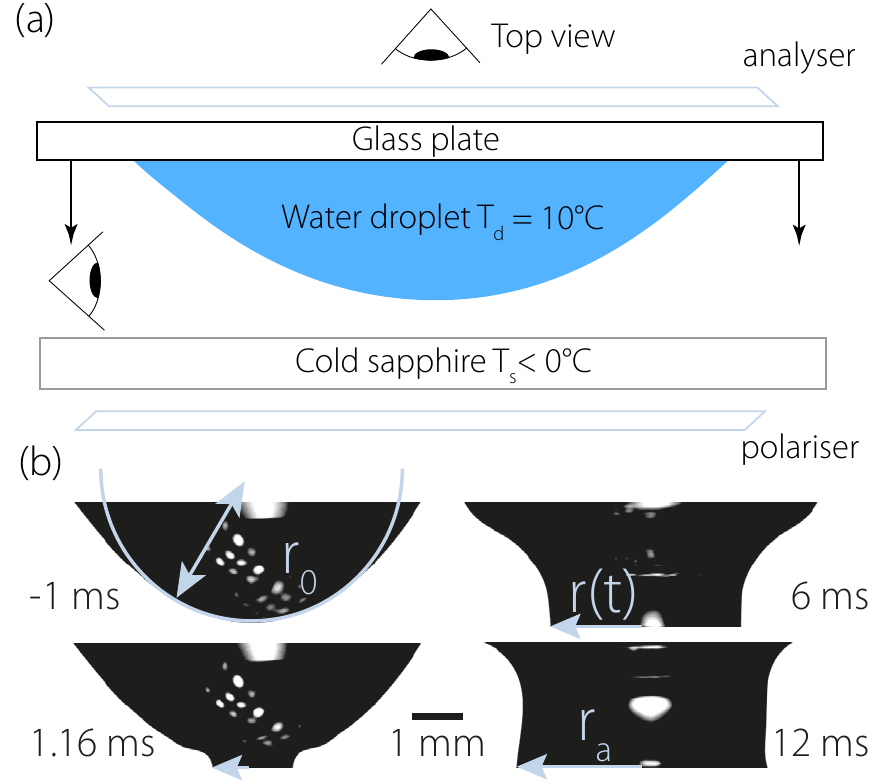}
\caption{(a) Schematic side-view of the experiment. 
(b) Temporal sequence of a droplet spreading on sapphire cooled at $T_{\rm s} = -3.2^{\circ}\, \rm C$ observed from the side. The contact line speed decreases with time until it gets pinned by solidification as seen in the final picture. 
}
\label{Manip}
\end{figure}

In the experiments, a droplet of initial tip radius $r_{\rm 0} = 3.1 \pm 0.23 \, \mathrm{mm}$
is brought in contact with a cold sapphire of temperature $T_{\rm s}$, thermal diffusivity $D_{\rm s}=11.5 \, \mathrm{mm^2.s^{-1}}$, and thickness $5 \, \mathrm{mm}$, cooled using liquid nitrogen. The drop-sapphire contact angle was measured as $\theta \approx 61 ^\circ$. The droplet is attached below a flat glass plate (see Figure~\ref{Manip} (a)) which is then moved downward
at low velocity until the drop touches the cold sapphire. As the spreading lasts less than a second, the position and the temperature of the glass plate can be considered constant during the experiment.
To avoid frost formation, the setup is placed in a humidity controlled box with relative humidity $H_{\rm r} < 3 \%$. 
The droplet initial temperature $T_{\rm d}$ is equal to the box temperature and constant in all experiments, close to $10^{\circ} \mathrm{C}$. The spreading dynamics is recorded from two synchronised high speed cameras at 6,000 fps. One provides a side-view of the spreading, using a 205mm lens and the other one a top-view, using a x12 Navitar lens. 
The top-view pictures are recorded under cross-polarised light to directly observe the crystal growth.


Figure \ref{Manip} (b) shows a side view recorded time sequence of a droplet spreading on the cold substrate.
After contact with the substrate, the contact line advances at early times symmetrically and ultimately the spreading stops, with an arrest radius $r_{\rm a}$. The spreading typically lasts few tenths of millisecond.

\begin{figure}[t]
    \centering
    \includegraphics[width=0.49\textwidth]{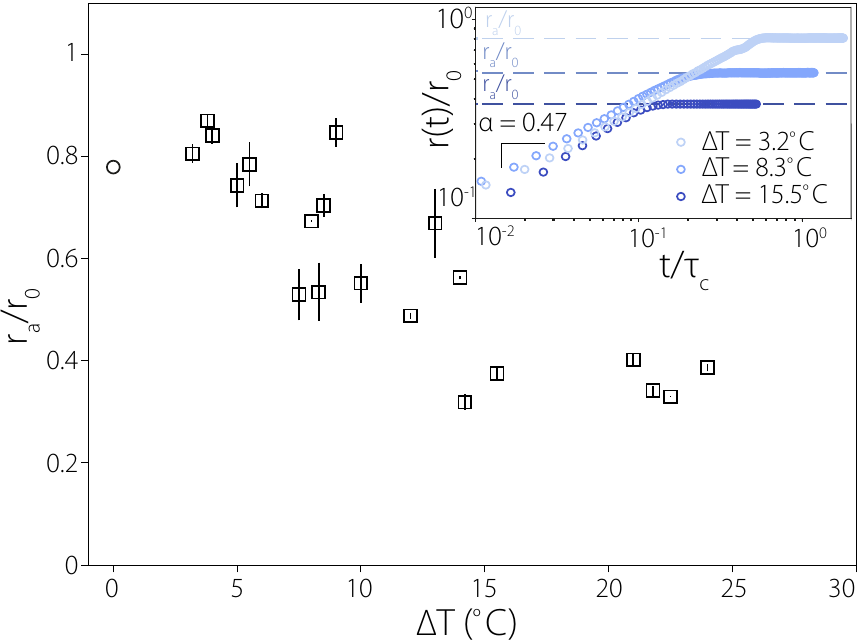}
    \caption{Rescaled arrest radius $r_{\rm a}/r_{\rm 0}$ as a function of the undercooling $\Delta T$. Error bars are computed based on spreading asymmetry. 
    The circular marker represents the isothermal case.
    The inset shows a log-log temporal evolution of the rescaled spreading radius. Before pinning, all curves superimpose and the coefficients fitted from Eq \ref{spreading} are ${C} = 1.15  \pm 0.06$ and $\alpha = 0.47 \pm 0.05$. The arrest diameter $r_{\rm a}/r_{\rm 0}$, in dotted lines (\protect\rule[0.5ex]{0.1cm}{0.2mm}\hspace{0.1cm}\protect\rule[0.5ex]{0.1cm}{0.2mm}\hspace{0.1cm}\protect\rule[0.5ex]{0.1cm}{0.2mm}) is determined at the time where the curve departs from the power-law behaviour.}
    \label{D_arret}
\end{figure}

Figure \ref{D_arret} displays the rescaled arrest radius $r_{\rm a}/r_{\rm 0}$ plotted against the undercooling $\Delta T = T_{\rm  m}- T_{\rm s}$ where $T_{\rm m} =0^{\circ} \rm{C}$ is the melting temperature of the ice. It shows a clear decrease of the maximum spreading radius as $\Delta T$ increases. 
For some experiments, the arrest radius is not the same on the left and the right sides of the image, an effect represented with the error-bars on the graph.
Noticeably, the arrest radius decreases by a factor 3 over the range of temperatures explored.
The inset of Figure~\ref{D_arret} displays, in a log-log scale, the evolution of the rescaled radius $r(t)/r_{\rm 0}$, with the time normalized by the inertial-capillary time $\tau_{\rm c}= \sqrt{\frac{\rho \, r_{\rm 0}^{3}}{\gamma}} \approx 21\, \rm{ms}$, for three undercoolings. It confirms that the substrate temperature has a strong effect on the arrest radius (see horizontal dashed lines $r_{\rm a}/r_{\rm 0}$), and shows that the spreading dynamics itself is neither affected by $\Delta T$ nor by the solidification \cite{de2017contact}.

In the first stage of the drop spreading, inertia resists to the liquid motion and the evolution of the droplet radius follows :
\begin{equation}
\frac{r(t)}{r_{\rm 0}} = C \left(\frac{t}{\tau_{\rm c}}\right)^{\alpha}
\label{spreading}
\end{equation} 

where the theoretical prediction gives $\alpha=1/2$ \cite{biance2004first}, valid for both wetting and partially wetting substrates at short times \cite{winkels2012initial}.

The inset of Figure~\ref{D_arret} shows that (i) the spreading dynamics observed on cooled substrates can be well modelled by this law until the time of pinning, (ii) the fitted coefficients $C$ and $\alpha$ are independent of the substrate temperature, and (iii) the exponent $\alpha = 0.47\pm 0.05$ agrees with the theoretical prediction (see also Supplementary Materials). These results are in agreement with former studies realised with other liquids and smaller temperature range \cite{de2017contact, koldeweij2021initial}.

\begin{figure*}
    \centering
    \includegraphics[width=\textwidth]{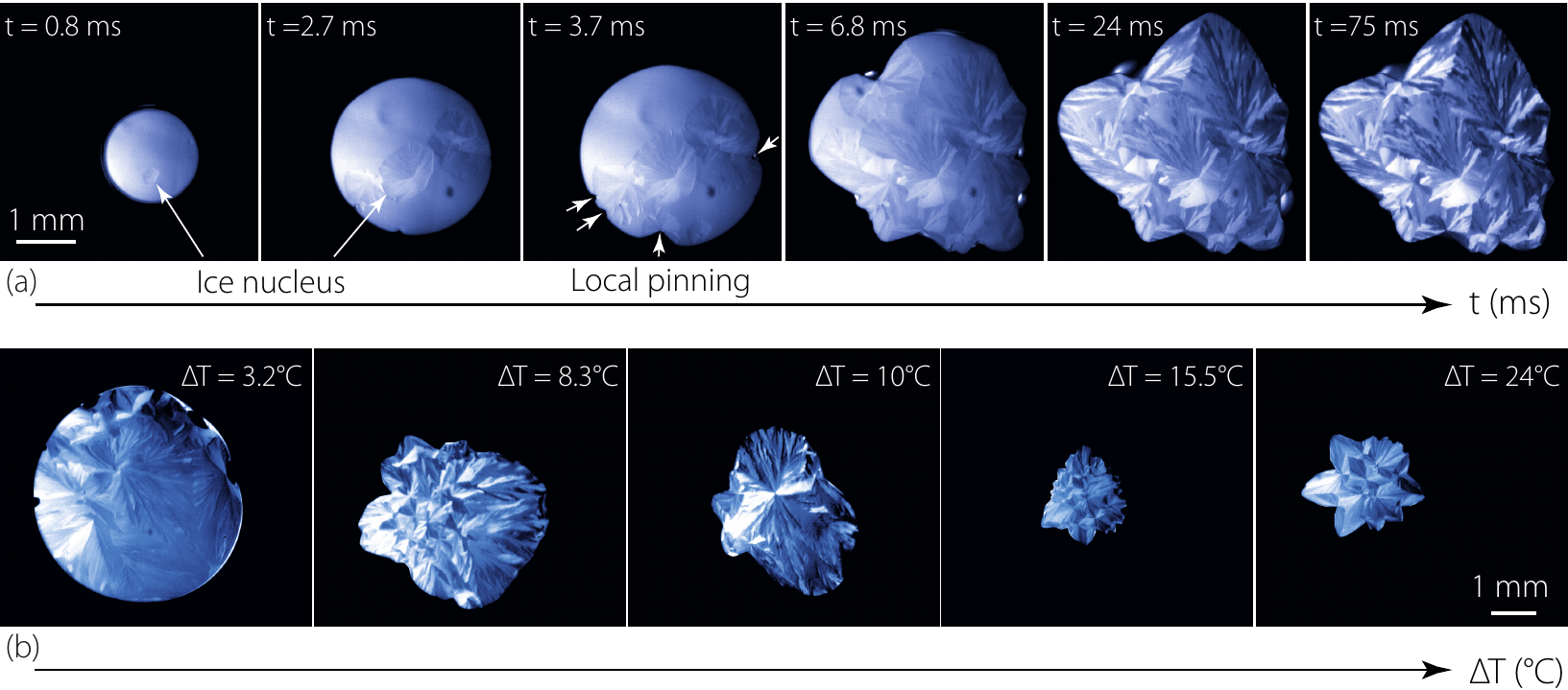}
    \caption{(a) Temporal evolution of the spreading at $-13 ^{\circ} \mathrm{C}$. Nucleation is designated by white arrows. Local pinning events are emphasised with small arrows at $t=3.7$ ms. (b) Final contact line shapes for different temperatures.}
\label{dessous}
\end{figure*}

To understand the mechanism responsible for the arrest of the contact line and its dependence with the substrate undercooling, we performed direct visualisations of the ice growth and its interaction with the spreading front.
Figure \ref{dessous} shows top views of the experiment under cross-polarised light, enabling a clear visualisation of the crystal growth, as seen on the sequence of Figure~\ref{dessous}~(a), and on the final shape of the deposits (Figure~\ref{dessous}~(b)). 
The temporal evolution can be divided into four phases. 
First, the drop spreads without ice nucleation: the contact line is circular (t$<$0.8 ms). 
Then, nucleation happens within the wetted disc (t=0.8 ms), and nuclei grow without affecting the contact line shape nor the dynamics (t$<$3.7 ms).
Third, the contact line gets locally pinned (t=3.7 ms), either by nucleation of crystals very close to it, as it is the case at the lower part of the drop, or by growing crystals catching it, as seen on the right of the drop between 2.7 and 3.7 ms. Here it seems that pinning occurs when the growing crystals catch the contact line.
Finally, the contact line is entirely stopped when pinned on its full perimeter by the crystals (t=24 ms). Then, the ice grows perpendicular to the substrate in the droplet bulk.

The final contact line shapes of the deposits, displayed on Figure~\ref{dessous}~(b), show a global decrease of the frozen drop radius with lower substrate temperatures, as seen on Figure~\ref{D_arret}. These pictures also enable to see clearly the crystal shape when the drop stops spreading.
In Figure~\ref{D_arret}, $r_\mathrm{a}$ is the mean radius determined from the side views, and larger error-bars at intermediate undercoolings are a consequence of the symmetry breaking in the contact line shape. Other definitions for the arrest radius have also been tested, such as the equivalent radius determined from the area of the final deposit, or the radius of the first pinned point of the contact line, showing however no qualitative change in the results (see Supplementary Materials).


From the time-sequences of Figure~\ref{dessous}~(a), the crystal-growth radial velocity $V_{\rm c}$ can be extracted for different temperatures of the substrate.
$V_{\rm c}$ is observed to be constant with time as shown in the Supplementary Materials. In a typical experiment, around 10 crystals are observed and their velocity is the same with small variations that can be attributed to thermal fluctuations within the water or the substrate. Thus, for each temperature, a single crystal-growth radial velocity $V_{\rm c}$ is measured with the associated relative error.

Similarly, the arrest velocity $V_{\rm a}$, the velocity of the contact line just before stopping, can be determined from the side views, taking the derivative of Eq. (\ref{spreading}) at the time of arrest $t_{\rm a}$, and using the fitted values of $C$ and $\alpha$
for each experiment: 

\begin{equation}
    V_{\rm a} = \frac{r_{\rm 0}C\alpha}{\tau_{\rm c}^{\alpha}} \, t_{\rm a}^{\alpha - 1} =\frac{r_{\rm 0}C\alpha}{\tau_{\rm c}} \, \left(\frac{r_{\rm a}}{C r_{\rm 0}}\right)^{\frac{\alpha - 1}{\alpha}}.
    \label{eq:Va}
\end{equation}

\begin{figure}[h]
    \centering
    \includegraphics[width=0.49\textwidth]{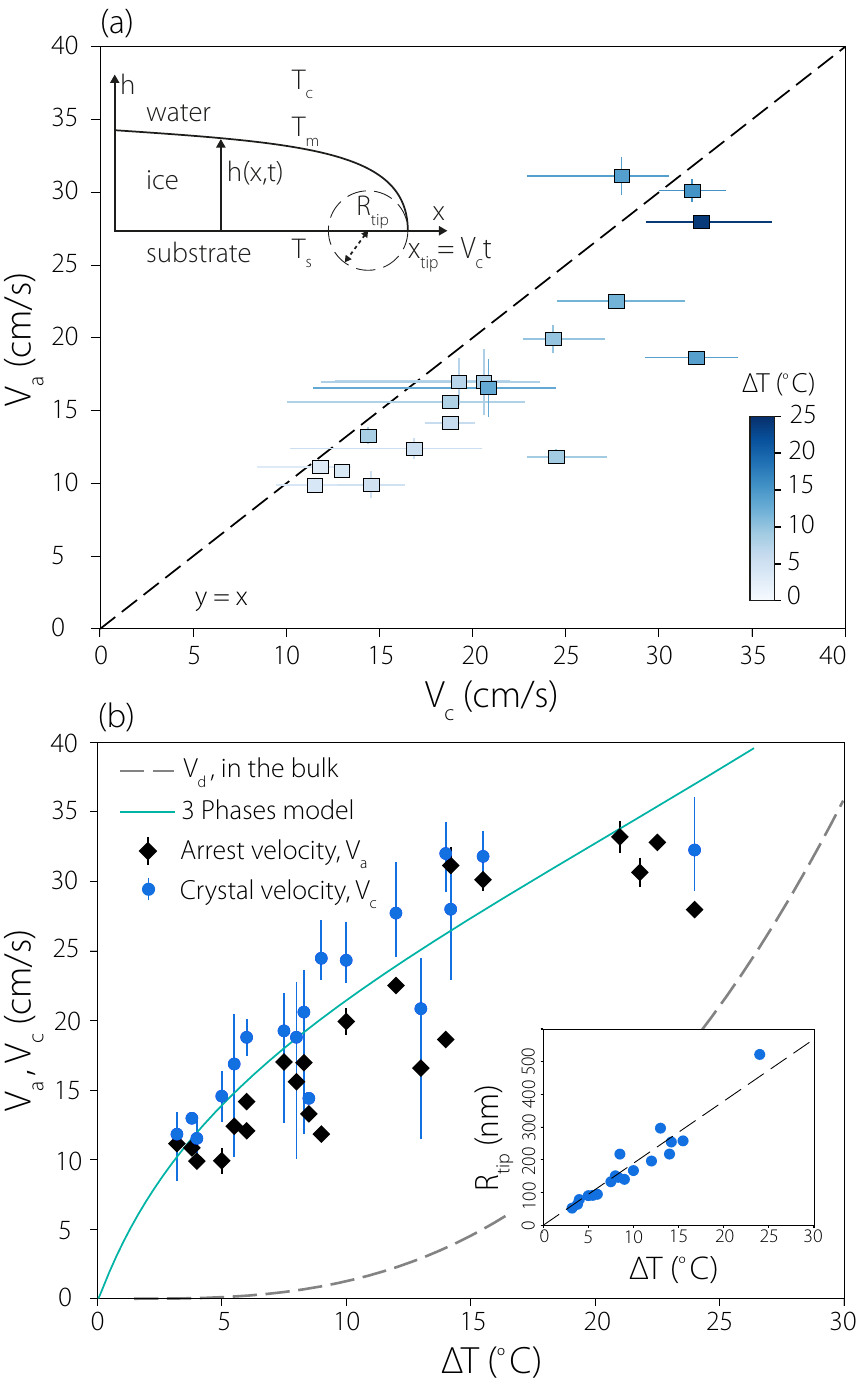}
    \caption{(a) Arrest velocity against crystal growth velocity. Colours indicate the surface temperature. The inset shows a sketch of the model of ice crystal growth near a substrate. (b) Arrest and crystal growth velocities as a function of the substrate undercooling. The dashed line represents the dentritic growth velocity and the plain one the model we developed. The inset shows that the tip radius of the crystal parabola increases linearly with $\Delta T$.} 
    \label{Arrest_velocity}
\end{figure}

As illustrated in the inset of Figure \ref{D_arret}, the transient time from spreading to arrest is very short, allowing a clear definition for the time of arrest $t_{\rm a}$ and its corresponding arrest velocity. The remaining uncertainty on $t_{\rm a}$ leads to error bars smaller than the ones induced by the asymmetry of the arrest.
In Figure \ref{Arrest_velocity} (a), the arrest velocity $V_{\rm a}$ is plotted as a function of the crystal growth velocity $V_{\rm c}$ for a large range of temperatures shown with a color bar. The dashed line represents the line $V_{\rm a}=V_{\rm c}$ and it is striking to notice that the arrest velocity
is always roughly equal to the crystal growth velocity.
This confirms our hypothesis on the mechanism responsible for the arrest of a contact line catched up by ice crystals: the contact line has to slow down to the crystal velocity to get caught by the crystal. Strictly speaking, this provides an upper limit for the arrest velocity. Hence,  the pinning occurs when $V_{\rm a} \lesssim V_{\rm c}$. Consequently, we observe smaller deposits when $\Delta T$ increases (Figure \ref{dessous}(a)): the contact line is caught earlier by the crystal as the crystal velocity increases with the undercooling.

Noticeably, in this scenario, the nucleation process plays a small role on the arrest criterion but could explain the small deviations observed, namely that the arrest velocity always seems to be slightly smaller than the crystal one (Figure~\ref{Arrest_velocity}~(a)). Once the velocity criterion is met, a "lag time" would still be necessary so that the contact line can be caught up by the crystals. Qualitatively, we often see that the contact line is first arrested by crystals nucleating very close to it and hence catching it quickly when the velocity criterion is met. This 
lag becomes negligible
at high undercoolings, where nucleation is very dense and uniformly distributed on the surface \cite{koldeweij2021initial, kant2020fast}. This could explain the small influence of the nucleation rate on the arrest mechanism.

Thanks to the visualisations, we can now discuss the measured crystal growth velocities.
As can be seen on Figure~\ref{dessous}~(a), part of the liquid is not immediately solidified after contact, and subsequently gets cooled  by the substrate. 
We can assume that the liquid temperature close to the substrate is the so-called contact temperature that arises when two bodies of different temperature are suddenly in contact $T_{\rm c} = T_{\rm s} + (T_{\rm d} - T_{\rm s})(1+ e_{\rm s}/e_{\rm w})$, where $e_{\rm s}$ and $e_{\rm w}$ are the effusivities of the substrate and water respectively, $e_{\rm k} = \sqrt{\lambda_{\rm k}\rho_{\rm k}C_{\rm p,k}}$ \cite{carslaw1959conduction}. 
In such configuration, the temperature in the water varies on a typical diffusive lengthscale $\delta \approx \sqrt{D_w \tau_c} \approx 10^{-5}$~m, with $D_w$ the heat diffusion coefficient in water. $\delta$ is thus far larger than the characteristic size of an ice nucleus (few nm) and of the tip radius of an ice dendrite growing in water at supercooled temperature $T_c$  (few $\mu$m) for our range of substrate temperatures \cite{shibkov2005crossover}. This allows us to approximate the temperature of the water surrounding the ice crystals as $T_{\rm c}$.

When ice grows in supercooled water, the ice formation occurs through a series of nucleation and growth of crystals \cite{libbrecht2017physical}. In the liquid bulk, the growth velocity is constant \cite{langer1980instabilities} but the mechanism that sets the velocity is not completely understood. Indeed, at large undercoolings, despite a good agreement between the experimental data available, theoretical models hardly catch the growth dynamics \cite{wang2019dendritic, shibkov2003morphology}.
More precisely, the work of Shibkov \textit{et al.} \cite{shibkov2003morphology} on ice growing freely in an infinite supercooled bulk provides a strong set of experimental data, and fitting a power-law over their whole experimental range gives $V_{\rm d} = K (T_{\rm m} - T_{\rm w})^\gamma$, $T_{\rm w}$ being the liquid undercooling, with $ K=4.83 \times 10^{-5} $~S.I and $\gamma= 2.78$ for the growth velocity, consistent with other studies \cite{glicksman1994dendritic,herbaut2019liquid}. This experimental fit is shown with the grey dashed line on Figure~\ref{Arrest_velocity}~(b) using $T_{\rm w}=T_{\rm c}$.
In this Figure, our velocities $V_{\rm a}$ (black diamonds) and $V_{\rm c}$ (blue dots) are represented as a function of $\Delta T$.
Our crystal growth velocities are systematically higher than the above power-law accounting for the ice growth velocity in liquid bulk.
In fact, several differences can be pointed out between these two experimental configurations, among which the heterogeneity of the water temperature, the motion in the surrounding fluid (however the two effects should lead to a decrease of the Shibkov prediction), and finally the presence of a cold substrate in the present case. 

The effect of the latter was recently explored \cite{schremb2017ice,kong2015theory} 
\cite{campbell2022dynamics} 
showing a strong coupling between the advancing crystal and the substrate. Indeed, latent heat due to solidification is also transferred to the substrate and in the case of a very effusive substrate ($e_{\rm s}>e_{\rm i}$) a higher solidification rate was clearly observed experimentally. 
To rationalize our experimental observations, we propose to enrich the model originally proposed by Schremb \textit{et al.} \cite{schremb2017ice}, which fails to reproduce our data on the whole range of investigated temperatures.
Following their approach, we assume that the vertical growth of the ice is well accounted by the 1D solidification problem, as illustrated on the inset of figure \ref{Arrest_velocity} (a). However, here, we take into account the thermal diffusion in all three phases (water, ice and substrate), and obtain that the growth of the ice front obeys a square-root law in time $h = \sqrt{D_{\rm eff}\Delta t}$ where $\Delta t$ is the time of solidification. $D_{\rm eff}$ depends on the substrate and water temperatures and is deduced from the self similar solution of the coupled heat equations for the three phases (see Sup. Mat.), generalising previous results~\cite{thievenaz2019solidification,thievenaz2020retraction}.
Using this square-root law solution and knowing that the crystal growth radial velocity is constant, we obtain: $h(x,t)=\sqrt{D_{\rm eff}( t - \frac{x}{V_{\rm c}})}=\sqrt{D_{\rm eff} \frac{x_\mathrm{tip}-x}{V_{\rm c}}}$.  
Subsequently, the shape of the crystal is a parabola and its tip radius is expressed as $R_{\rm tip}=\frac{D_\mathrm{eff}}{2\, V_{\rm c}}$. 

Measuring $V_{\rm c}$ and calculating the theoretical value of $D_\mathrm{eff}$, we can compute $R_{\rm tip}$ for each of our experiments, plotted in the inset of Figure~\ref{Arrest_velocity}~(b) as a function of the undercooling. By contrast with former studies, where $R_{\rm tip}$ was found constant \cite{schremb2017ice,kong2015theory} or with dendritic growth where the tip radius decreases with the undercooling, we obtain a clear linear dependence $R_{\rm tip}=k\Delta T$, with $k=2. 10^\mathrm{-8} \, \rm m/K$. Using this fitted law for $R_{\rm tip}$ in the formula above relating the tip radius and $V_c$, we obtain $V_c=\frac{D_{\rm eff}}{2k \Delta T}$ corresponding to the green plain curve on Figure~\ref{Arrest_velocity}~(b) which, by construction, shows a very good agreement with our experiments. The linear law for $R_{\rm tip}$, together with our refined model, allows a good understanding of our results over the whole temperature range.

In conclusion, when a drop spreads on a cold substrate, its decelerating contact line is eventually caught up and arrested by ice crystals growing at the substrate-water interface. The race between the spreading line and the solidification front determines an arrest criterion, qualitatively similar to the ones proposed in previous studies  \cite{de2017contact,herbaut2019liquid,koldeweij2021initial}. However, the new experiments proposed here, on water and at large undercoolings, provide a direct visualisation of the growing crystals, and the model reveals the role of the substrate thermal properties on the arrest. 
Finally, the physical mechanism leading to the pinning of the contact line when reached by the crystal still needs to be elucidated. Several theories have already been proposed
\cite{tavakoli2014spreading,schiaffino1997molten,herbaut2020criterion} and this phenomenon, which takes place at very small time and length scales, should be the object of future experimental investigations.

\begin{acknowledgments}

We thank the Direction G\'en\'erale de l'Armement (DGA) for financial support. \\

R. Grivet and A. Monier contributed equally.

\end{acknowledgments}

\bibliographystyle{apsrev4-2}
\bibliography{Biblio_MonierGrivet}

\end{document}